\documentclass[lettersize,journal]{IEEEtran}
\IEEEoverridecommandlockouts

\usepackage{amssymb}
\usepackage[cmex10]{amsmath}
\usepackage{stfloats}
\usepackage{graphicx}
\usepackage{subfigure}
\usepackage{tabularx}
\usepackage{epsfig,epsf,color,balance,cite}
\usepackage{verbatim}
\usepackage{url}
\usepackage{bm}
\usepackage{booktabs}

\usepackage{amsmath}

\usepackage{algorithm}
\usepackage{algorithmic}

\usepackage{color}
\definecolor{myc1}{rgb}{0,0,0}

\begin{document}

\title{A Joint Communication and Computation Design for Semantic Wireless Communication with Probability Graph}

\author{Zhouxiang Zhao, 
        Zhaohui Yang, 
        Xu Gan, 
        Quoc-Viet Pham, 
        Chongwen Huang, \\
        Wei Xu, 
        and Zhaoyang Zhang
\thanks{Z. Zhao, Z. Yang, X. Gan, C. Huang, and Z. Zhang are with College of Information Science and Electronic Engineering, Zhejiang University, and also with Zhejiang Provincial Key Laboratory of Info. Proc., Commun. \& Netw. (IPCAN), Hangzhou, China (e-mails: \{zhouxiangzhao, yang\_zhaohui, gan\_xu, chongwenhuang, ning\_ming\}@zju.edu.cn).}
\thanks{Q. Pham is with School of Computer Science and Statistics, Trinity College Dublin, Ireland (e-mail: viet.pham@tcd.ie).}
\thanks{W. Xu is with National Mobile Communications Research Laboratory, Southeast University, Nanjing, China (e-mail: wxu@seu.edu.cn).}
}

\maketitle

\begin{abstract}
In this paper, we delve into the challenge of optimizing joint communication and computation for semantic communication over wireless networks using a probability graph framework. In the considered model, the base station (BS) extracts the small-sized compressed semantic information through removing redundant messages based on the stored knowledge base. Specifically, the knowledge base is encapsulated in a probability graph that encapsulates statistical relations. At the user side, the compressed information is accurately deduced using the same probability graph employed by the BS. While this approach introduces an additional computational overhead for semantic information extraction, it significantly curtails communication resource consumption by transmitting concise data. We derive both communication and computation cost models based on the inference process of the probability graph. Building upon these models, we introduce a joint communication and computation resource allocation problem aimed at minimizing the overall energy consumption of the network, while accounting for latency, power, and semantic constraints. To address this problem, we obtain a closed-form solution for transmission power under a fixed semantic compression ratio. Subsequently, we propose an efficient linear search-based algorithm to attain the optimal solution for the considered problem with low computational complexity. Simulation results underscore the effectiveness of our proposed system, showcasing notable improvements compared to conventional non-semantic schemes.
\end{abstract}

\begin{IEEEkeywords}
Semantic communication, knowledge graph, probability graph, joint communication and computation.
\end{IEEEkeywords}
\IEEEpeerreviewmaketitle

\section{Introduction}
\IEEEPARstart{O}{ver} the preceding decades, the evolution of mobile communication technologies has played a pivotal role in advancing human society. Shannon's introduction of information theory in the 1940s focused on quantifying the maximum achievable data transmission rate through a communication channel \cite{shannon1948mathematical}. Subsequently, contemporary communication systems have predominantly been designed with an emphasis on metrics related to transmission rates \cite{larsson2014massive,8741198}. However, the escalating demand for intelligent applications in wireless communication necessitates a shift from traditional architectures, solely prioritizing high transmission rates, to novel architectures geared towards task efficiency \cite{xu2023edge,chen2023big,10283760}. In the realm of future wireless communications, mobile information networks will confront the challenge of accommodating tasks with low latency and high efficiency \cite{gunduz2022beyond}. This evolving landscape has given rise to a new communication paradigm known as \emph{semantic communication}. Unlike Shannon's theory, which did not attribute significance to the semantic information of data, deeming it largely irrelevant to communication, the concept of semantic communication was introduced by Warren Weaver \cite{weaver1953recent}. Weaver's three-layer communication framework delineates the technical layer, concerned with traditional communication and accurate transmission of symbols; the semantic layer, focused on conveying the meaning of communication symbols; and the effectiveness layer, concentrated on how the received meaning influences the receiver's behavior. Semantic communication represents an innovative architecture capable of integrating user needs and information meaning into the communication process \cite{chaccour2022less,10233741}. Unlike traditional communication systems dedicated solely to the reliable transmission of bit streams without knowledge of the message's meaning or exchange goals, semantic communication systems prioritize conveying meaning in the message \cite{weng2021semantic_icc,10333452}. Consequently, they can significantly reduce the required transmission channel bandwidth \cite{lan2021semantic}. In the context of future mobile communication systems, semantic communication is anticipated to be a pivotal technology for the development of sixth-generation (6G) networks \cite{qin2021semantic,8476247}.

In recent studies, various investigations have delved into diverse aspects of semantic communication. In \cite{xie2021deep}, the authors proposed a semantic communication system for text transmission based on deep learning (DL). They introduced a novel metric, sentence similarity, to assess the effectiveness of semantic communication. Another contribution in \cite{weng2021semantic} focused on a DL-enabled semantic communication system tailored for speech signals. The incorporation of an attention mechanism, utilizing a squeeze-and-excitation network, aimed to enhance the accuracy of speech signal recovery, particularly for crucial information. Additionally, work \cite{tong2021federated} addressed the transmission of audio semantic information capturing contextual features of audio signals. They introduced a wave-to-vector architecture-based autoencoder, comprising convolutional neural networks, to extract semantic information from audio signals. Moreover, work \cite{xie2020lite} presented a lite distributed semantic communication system named L-DeepSC, designed for text transmission with low complexity. Operating at the semantic level, this system facilitates data transmission from Internet of Things (IoT) devices to the cloud/edge, thereby improving transmission efficiency. In the domain of semantic-oriented speech transmission, work \cite{han2022semantic} proposed an end-to-end DL-based transceiver that extracts and encodes semantic information from input speech spectrums at the transmitter, delivering corresponding transcriptions from the decoded semantic information at the receiver. Furthermore, work \cite{peng2022robust} explored the mechanism of semantic noise and introduced a robust DL-enabled semantic communication system. Leveraging a calibrated self-attention mechanism and adversarial training, this system addresses semantic noise challenges. However, it is noteworthy that the works mentioned above \cite{xie2021deep, weng2021semantic, tong2021federated, xie2020lite, han2022semantic, peng2022robust} did not account for computation energy consumption in their assessments of semantic communication systems. Previous researches \cite{10032275, 9763856} have demonstrated that efficient resource allocation plays a crucial role in significantly improving the performance of semantic communication.

\begin{figure}[t]
\centering
\includegraphics[width=3.4in]{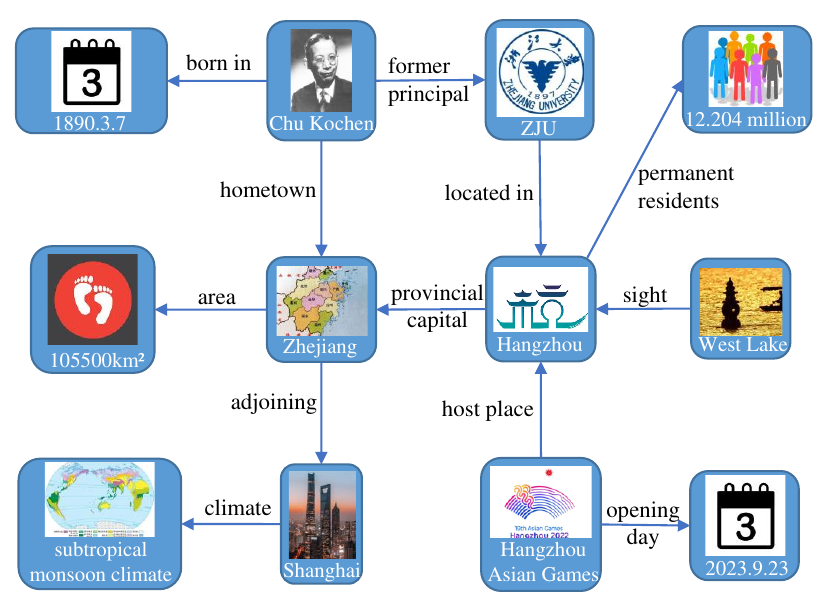}
\caption{An example of knowledge graph.}
\label{fig1}
\end{figure}

\begin{figure*}[t]
\centering
\includegraphics[width=6.5in]{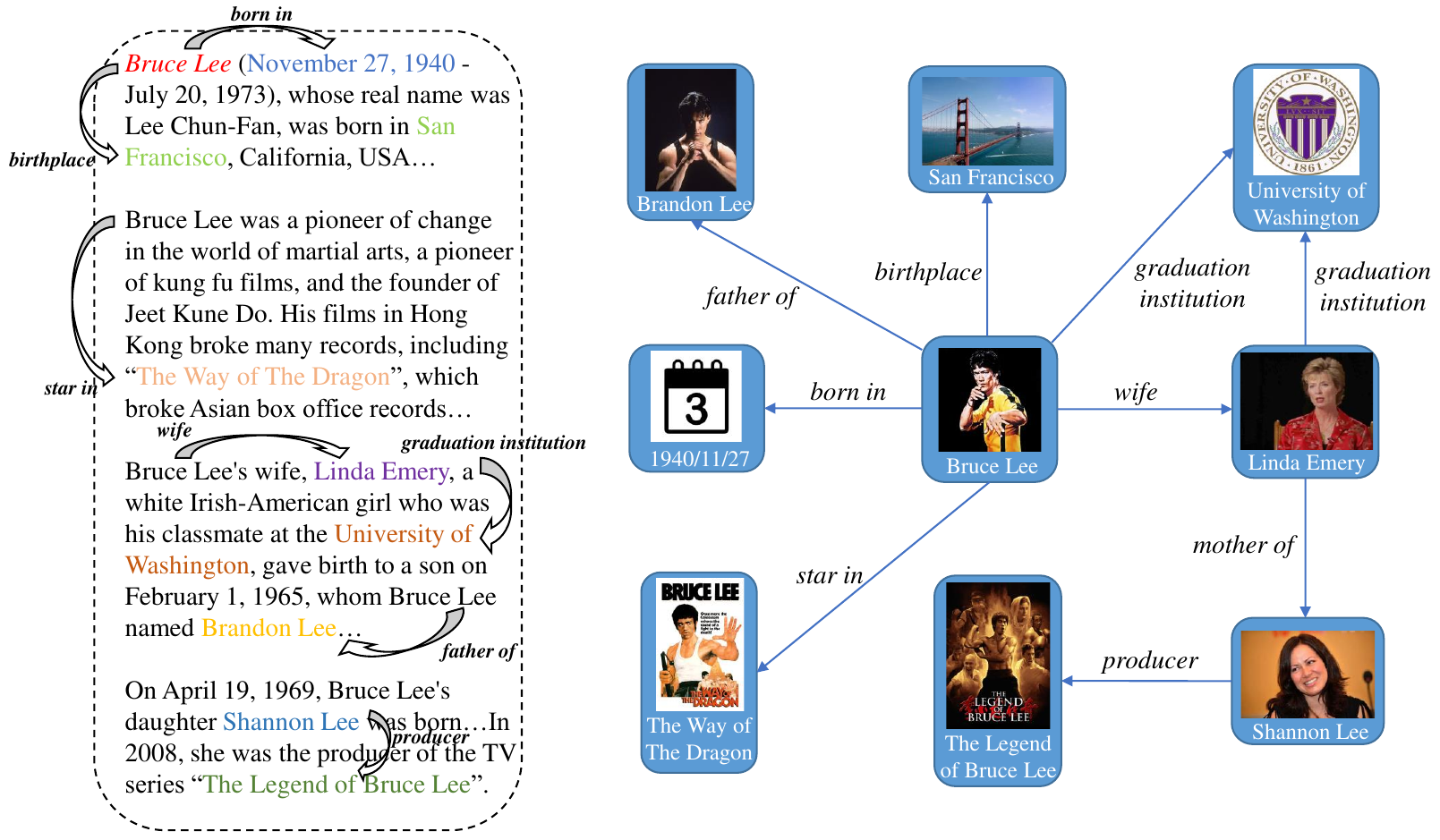}
\caption{The process of extracting semantic triples from text data.}
\label{fig2}
\end{figure*}

Despite the progress made in semantic communication research, the field is still in its early stages, facing challenges that impede its development, notably the absence of unified standards. One such challenge is semantic representation \cite{griffiths2019probabilistic}, aimed at finding suitable ways to represent semantic information. Existing studies \cite{al2021hybrid,7998610} utilize transmitted content features for semantic representation, but this approach does not align with the logic of human language and lacks validation through interaction with human understanding. To address these challenges, this paper adopts knowledge graphs \cite{9416312,7358050} as a means of representing the semantic information of text data. Knowledge graphs, structured forms of knowledge with high information density, are pivotal technical tools for extracting semantic information \cite{bonatti2019knowledge}. Comprising entities and relations, where entities represent real-world objects and abstract concepts, and relations capture concrete relationships between entities, knowledge graphs employ triples in the form of (\emph{head, relation, tail}) to represent knowledge, e.g., (\emph{banana, a kind of, fruit}). Named entity recognition (NER) \cite{9039685} and relation extraction (RE) \cite{9675156} are crucial techniques for constructing structured knowledge graphs from unstructured textual data. NER identifies entities (people, places, organizations, etc.), while RE detects semantic relations between these entities. The interpretability of knowledge graphs has spurred research on the relationship between knowledge graphs and semantics. Notably, \cite{jaradeh2019open} posited that knowledge graphs represent the next-generation infrastructure for semantic scholarly knowledge, \cite{atef2021predicting} suggested the potential use of knowledge graphs in predicting semantic categories, and \cite{9838470} integrated knowledge graphs with semantic communication to enhance communication credibility. Despite the effectiveness of knowledge graphs in representing semantics, existing knowledge graph-based semantic communication works \cite{9838470,jiang2022reliable,10061867} have not established a knowledge base to further reduce communication load in semantic wireless communication networks. While semantic communication techniques offer the advantage of reducing transmitted data size, they introduce additional computational overhead. Consequently, the allocation of resources for both communication and computation remains an underexplored area. To the best of our knowledge, this work is the first to utilize a probability graph as the knowledge base in semantic communication, considering both communication and computation resource allocation.

The primary contribution of this paper lies in the development of an innovative text information extraction system based on probability graph, which takes into consideration both the communication and computation energy consumption within the semantic communication network. The key contributions can be succinctly summarized as follows:

\begin{itemize}
\item We introduce a novel knowledge graph compression method based on a probability graph shared by both the base station (BS) and the user. The probability graph, derived from knowledge graphs, extends the traditional triple to a quadruple by incorporating the dimension of relation probability. Leveraging this shared probability graph allows the omission of redundant information in the knowledge graph associated with the text, leading to a reduction in transmitted data.
\item We provide a computation energy cost model for inferring the reduced information based on the probability graph. This model serves as the foundation for addressing the joint communication and computation resource allocation problem in the semantic communication system. Our approach involves the simultaneous optimization of transmit power and compression ratio. The formulated optimization problem aims to minimize the overall communication and computation energy consumption of the network, while adhering to constraints related to latency, transmit power, and semantic considerations.
\item To solve the formulated problem, we develop an efficient algorithm that computes the closed-form power allocation for each omitted relation number. The proposed algorithm is validated through numerical analysis, demonstrating its effectiveness in optimizing resource allocation in the semantic communication system.
\end{itemize}

The rest of this paper is organized as follows. The system model and problem formulation are described in Section \uppercase\expandafter{\romannumeral2}. The algorithm design is presented in Section \uppercase\expandafter{\romannumeral3}. Simulation results are analyzed in Section \uppercase\expandafter{\romannumeral4}. Future research directions are introduced in Section \uppercase\expandafter{\romannumeral5}. Conclusions are drawn in Section \uppercase\expandafter{\romannumeral6}.

\section{System Model and Problem Formulation}

\begin{figure}[t]
\centering
\includegraphics[width=3.4in]{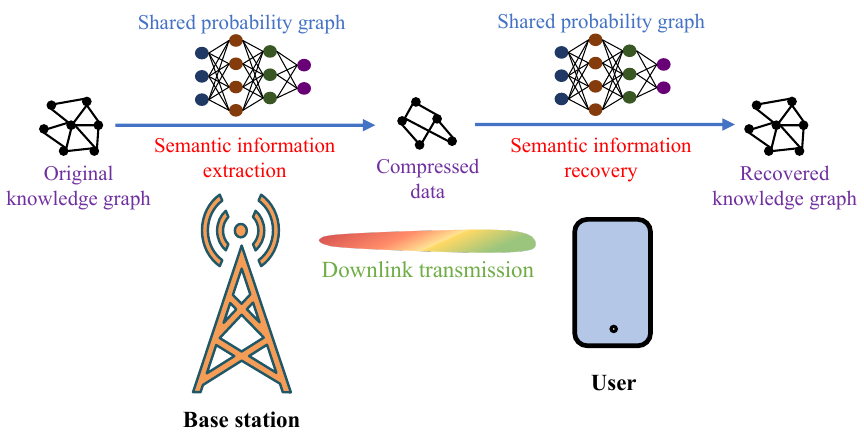}
\caption{An illustration of the considered semantic communication system.}
\label{fig3}
\end{figure}

Consider a semantic wireless communication system with one BS and one user, as illustrated in Fig. \ref{fig3}. The BS and the user share common background knowledge, which is characterized by the probability graph. The BS needs to transmit a substantial amount of text data with similar topics to the user while accounting for limited wireless communication resources. To achieve this goal, the BS extracts semantic information from the raw text data and compresses the semantic information (represented by a knowledge graph) using a probability graph shared with the user before transmitting it through the wireless channel. Subsequently, the user recovers the original knowledge graph based on the probability graph shared with the BS.

\subsection{Probability Graph Model}
\begin{figure}[t]
\centering
\includegraphics[width=3.4in]{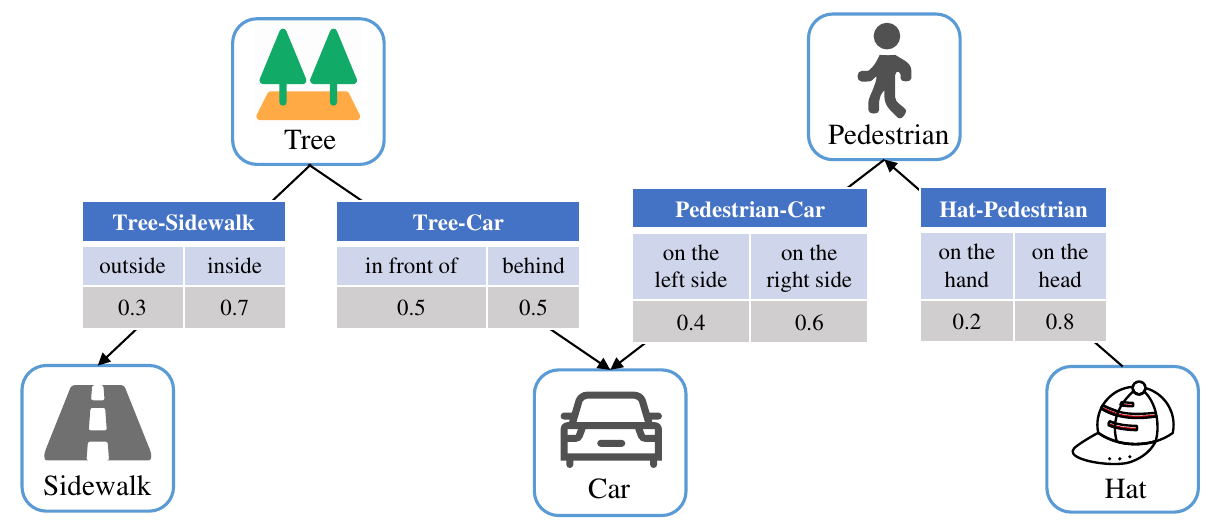}
\caption{Illustration of the probability graph considered in the semantic communication system.}
\label{fig4}
\end{figure}

The semantic information within a knowledge graph is conventionally expressed as triples in the form (\emph{head, relation, tail}). From a given piece of text data, multiple triples can be extracted, collectively characterizing a knowledge graph. In scenarios where multiple texts center around similar topics, certain triples may share common head and tail entities while differing in specific relations. For example, text data 1 might contain (\emph{Tree, in front of, Car}), while text data 2 might have (\emph{Tree, behind, Car}). In such cases, combining these triples could be more informative and efficient. Thus, the representation could be \{\emph{Tree}, [(\emph{in front of}, 0.5), (\emph{behind}, 0.5)], \emph{Car}\}, with the numeric values indicating the frequency of occurrence of each relation. When confronted with a substantial volume of text data sharing similar topics, their corresponding knowledge graphs can be amalgamated into a comprehensive probability graph using a similar method. In this context, the edges in the knowledge graph are not strictly determined relational edges but probabilistic edges, combining the frequency of occurrence of different relations, as illustrated in Fig. \ref{fig4}.

Before initiating wireless transmission, the BS constructs a probability graph based on a significant number of historical samples. Subsequently, the BS transmits this probability graph to the user as common background knowledge. This process occurs once, and wireless transmission commences afterward.

During the wireless transmission process, the BS transmits the required text data to the user. The sample dataset is formulated as
\begin{equation}\label{eq1}
\mathcal T = \left\{T_1,T_2, \cdots ,T_n, \cdots ,T_N\right\},
\end{equation}
where $N$ represents the total number of sample data, $T_n$ denotes the $n$-th sample data, and $\mathcal T$ denotes the sample dataset.

The knowledge graph extracted from each sample data $T_n$ is represented by
\begin{equation}\label{eq2}
G_n = \left\{\varepsilon^1_n,\varepsilon^2_n,\cdots,\varepsilon^m_n,\cdots,\varepsilon^M_n\right\},
\end{equation}
where $\varepsilon^m_n$ denotes the $m$-th triple in knowledge graph $G_n$, and $M$ denotes the total number of triples. The triple $\varepsilon^m_n$ can be expressed as
\begin{equation}\label{eq3}
\varepsilon^m_n = \left(h^m_n, r^m_n, t^m_n\right),
\end{equation}
where $h^m_n$ represents the head entity of triple $\varepsilon^m_n$, $t^m_n$ represents the tail entity of triple $\varepsilon^m_n$, and $r^m_n$ represents the relation between $h^m_n$ and $t^m_n$.

The probability graph shared by the BS and the user can be represented by
\begin{equation}\label{eq4}
\mathbb{G} = \left\{\delta_1, \delta_2, \cdots, \delta_s, \cdots, \delta_S\right\},
\end{equation}
where $\delta_s$ is the quadruple with relational probability, and $S$ is the total number of quadruples. Specifically, $\delta_s$ can be calculated as
\begin{equation}\label{eq5}
\delta_s=\left\{h_s,\left[\left(r_s^1, \mathcal{N}_s^1\right), \cdots,\left(r_s^i, \mathcal{N}_s^i\right), \cdots,\left(r_s^I, \mathcal{N}_s^I\right)\right], t_s\right\},
\end{equation}
where $h_s$ is the head entity of quadruple $\delta_s$, $t_s$ is the tail entity of quadruple $\delta_s$, $r_s^i$ is the $i$-th relation between $h_s$ and $t_s$, $I$ is the total number of different relations between $h_s$ and $t_s$, and $\mathcal{N}_s^i$ is the set of samples in which the triple $\left(h_s, r^i_s, t_s\right)$ holds. For instance, if $\left(h_s, r^i_s, t_s\right)$ holds in samples $T_1$, $T_4$, and $T_7$, then $\mathcal{N}_s^i = \left\{1,4,7\right\}$.

Building upon equations \eqref{eq4} and \eqref{eq5}, the construction of the probability graph shared by the BS and the user is successfully completed.

\subsection{Probability Graph-enabled Information Compression}
Once the set of quadruples has been obtained, it is possible to calculate the multidimensional conditional probability distribution. The methodology will be elaborated upon in the following.

If no prior information is available, the probability that the triple $\left(h_s, r^i_s, t_s\right)$ holds can be written as
\begin{equation}\label{eq6}
    p\left(h_s, r_s^i, t_s\right)=\frac{\operatorname{num}\left(\mathcal{N}_s^i\right)}{\sum_{i=1}^I \operatorname{num}\left(\mathcal{N}_s^i\right)},
\end{equation}
where $\operatorname{num}\left(\mathcal{N}_s^i\right)$ is the number of elements in the set $\mathcal{N}_s^i$.
If the triple $\left(h_s, r^i_s, t_s\right)$ is known to be true, then the conditional probability of the triple $\left(h_k, r^i_k, t_k\right)$ can be expressed as
\begin{equation}\label{eq7}
    p\left[\left(h_k, r_k^i, t_k\right) \mid\left(h_s, r_s^i, t_s\right)\right]=\frac{\operatorname{num}\left(\mathcal{N}_k^i \cap \mathcal{N}_s^i\right)}{\operatorname{num}\left[\mathcal{N}_s^i \cap\left(\bigcup_{i=1}^I \mathcal{N}_k^i\right)\right]},
\end{equation}
where $\mathcal{N}_k^i \cap \mathcal{N}_s^i$ is the intersection of set $\mathcal{N}_k^i$ and set $\mathcal{N}_s^i$, and $\bigcup_{i=1}^I \mathcal{N}_k^i$ is the union of set $\mathcal{N}_k^1$ to $\mathcal{N}_k^I$.

Using a similar approach, it is appropriate to derive the multidimensional conditional probability of the triple $\left(h_k, r^i_k, t_k\right)$ given that $N$ triples are known to be true:
\begin{equation}\label{eq8}
    p\left[\left(h_k, r_k^i, t_k\right) \mid \mathcal{C}_N\right]=\frac{\operatorname{num}\left[\mathcal{N}_k^i \cap\left(\cap_{\mathcal{N} \subset \mathcal{C}_N} \mathcal{N}\right)\right]}{\operatorname{num}\left[\left(\cap_{\mathcal{N} \subset \mathcal{C}_N} \mathcal{N}\right) \cap\left(\bigcup_{i=1}^I \mathcal{N}_k^i\right)\right]},
\end{equation}
where $\mathcal{C}_N$ is a set of conditions that indicates the truth of $N$ triples, and $\cap_{\mathcal{N}\subset \mathcal{C}_N}\mathcal{N}$ denotes the intersection of all sample sets that satisfy the condition set $\mathcal{C}_N$.

\begin{figure}[t]
\centering
\includegraphics[width=3in]{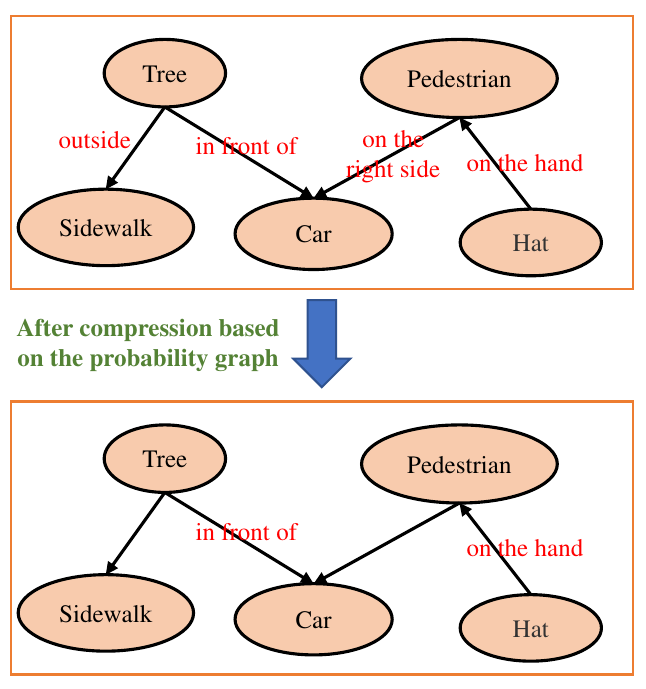}
\caption{An illustration of the information compression based on the probability graph.}
\label{fig5}
\end{figure}

When transmitting the knowledge graph that corresponds to a specific text data, the multidimensional conditional probability distribution described above can be used to remove semantic relations for the purpose of information compression, as depicted in Fig. \ref{fig5}. The compression process can be achieved using the following scheme.

Prior to transmitting the text data, the BS first applies natural language processing techniques to extract the corresponding knowledge graph denoted as $G$, which is expressed as
\begin{equation}\label{eq9}
    G = \left\{\varepsilon_1, \varepsilon_2, \cdots, \varepsilon_j, \cdots, \varepsilon_J\right\},
\end{equation}
where $\varepsilon_j$ is the $j$th triple in $G$, and $J$ is the total number of triples in $G$. Specifically, the triple $\varepsilon_j$ can be written in the following form:
\begin{equation}\label{eq10}
    \varepsilon_j = \left(h_j,r_j,t_j\right).
\end{equation}

Next, the knowledge graph $G$ is compared with the shared probability graph $\mathbb{G}$, and the comparison process can be divided into several rounds.

\begin{figure}[t]
\centering
\includegraphics[width=3.4in]{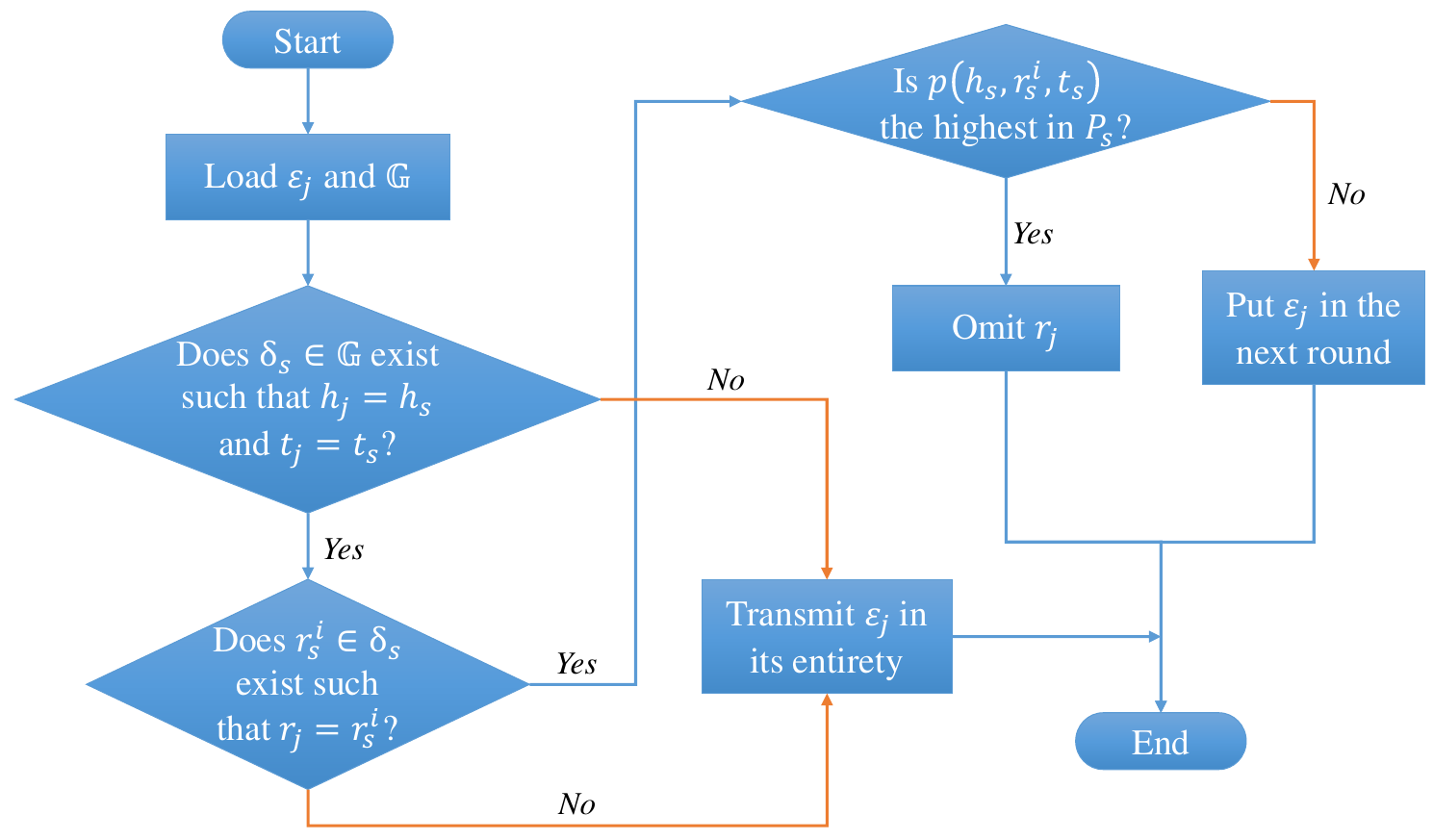}
\caption{The flowchart of the comparison process.}
\label{f}
\end{figure}

In the initial round, there is no prior information. For any $\varepsilon_j$, we need to examine if there exists $\delta_s$ in $\mathbb{G}$ where $h_j=h_s$ and $t_j=t_s$. If such a case is found, we proceed to investigate whether there exists a relation $r_s^i$ within the relation set of $\delta_s$ that matches $r_j$. If a match is identified, we further determine whether the corresponding relation probability $p\left(h_s,r_s^i,t_s \right)$ is the highest among the set of relation probabilities denoted by $P_s$. It is possible to represent $P_s$ as
\begin{equation}\label{eq11}
    P_s = \left\{p\left(h_s,r^1_s,t_s\right),\cdots,p\left(h_s,r^i_s,t_s\right),\cdots,p\left(h_s,r^I_s,t_s\right)\right\}.
\end{equation}
If $p\left(h_s,r_s^i,t_s \right)$ happens to be the largest among the relation probabilities in set $P_s$, we can exclude the relation $r_j$ from the triple $\varepsilon_j=\left( h_j,r_j,t_j \right)$ during transmission. In this case, only the head entity and tail entity need to be sent, reducing the transmitted data volume. However, if there is no $\delta_s$ in $\mathbb{G}$ satisfying $h_j=h_s$ and $t_j=t_s$, or if such a $\delta_s$ exists but there is no relation $r_s^i$ in its relation set that matches $r_j$, then the entire triple must be transmitted without omission. The flowchart of the comparison process is shown in Fig.~\ref{f}.

In the second round, one-dimensional conditional probabilities can be employed as prior information. As certain triples had their semantic relations omitted in the first round, these triples can serve as conditions for the second round's conditional probability search. The second round of comparison follows a similar process as the first round. We can denote the set of triples that were not omitted in the first round and do not necessarily require full transmission as
\begin{equation}\label{eq12}
    G^1 = \left\{\varepsilon_1,\varepsilon_2,\cdots,\varepsilon_a,\cdots,\varepsilon_A\right\}.
\end{equation}
The triples that can be inferred in the probability graph can be omitted to save the communication burden. The set of of omitted triples can be represented by
\begin{equation}\label{eq13}
    O^1 = \left\{\epsilon_1,\epsilon_2,\cdots,\epsilon_b,\cdots,\epsilon_B\right\}.
\end{equation}

For every $\varepsilon_a$ in $G^1$, we need to compute its conditional probability given each triple $\epsilon_b$ in $O^1$. Assume that $\varepsilon_a$ corresponds to $\delta_a = \left(h_a,r_a^i,t_a \right)$ in the shared probability graph $\mathbb{G}$. The conditional probability can be represented as $p\left[\left(h_a,r_a^i,t_a \right)\mid \epsilon_b \right]$, which can be denoted by $p^1_{a\mid b}$ for simplicity. Consequently, we can construct a probability matrix consisting of one-dimensional conditional probabilities\footnote{In the subsequent rounds, they will be multi-dimensional conditional probabilities.} for $\left(h_a,r_a^i,t_a \right)$ as follows:
\begin{equation}\label{eq14}
    P^1_{\delta_a} = \begin{bmatrix}
        p^1_{1 \mid 1} & p^1_{2 \mid 1} & \dots & p^1_{i \mid 1} & \dots & p^1_{I \mid 1}\\
        p^1_{1 \mid 2} & p^1_{2 \mid 2} & \dots & p^1_{i \mid 2} & \dots & p^1_{I \mid 2}\\
        \vdots         & \vdots         & \ddots& \vdots         & \ddots& \vdots        \\
        p^1_{1 \mid b} & p^1_{2 \mid b} & \dots & p^1_{i \mid b} & \dots & p^1_{I \mid b}\\
        \vdots         & \vdots         & \ddots& \vdots         & \ddots& \vdots        \\
        p^1_{1 \mid B} & p^1_{2 \mid B} & \dots & p^1_{i \mid B} & \dots & p^1_{I \mid B}\\
        \end{bmatrix},
\end{equation}
where $I$ is the total number of different relations between $h_a$ and $t_a$ in $\mathbb{G}$, and $B$ is the total number of elements in $O^1$.
In the probability matrix $P^1_{\delta_a}$, we begin with the first row and examine whether $p^1_{i \mid b}$ is the largest element within that row. If it is indeed the largest, we take note of the column number and exclude the relation $r^i_a$ when transmitting the corresponding triple. However, if $p^1_{i \mid b}$ is not the largest element in any of the rows, it indicates that the corresponding relation cannot be omitted during transmission.

We repeat the aforementioned procedure for every triple in $G^1$, aiming to identify the triples that can be omitted. This constitutes the first cycle in the second round. In the second cycle, there will be fewer remaining triples that have not been omitted. We create a new $G^1$ using these unomitted triples and form a new $O^1$ for the newly omitted triples. We continue this process until no new triples can be omitted, iterating through multiple cycles if necessary.

In the third round, we incorporate two-dimensional conditional probabilities as prior information. We derive the corresponding $G^2$ and $O^2$ based on the previous round's results. Subsequently, we calculate a three-dimensional probability matrix, denoted by $P^2$, which comprises two-dimensional conditional probabilities.

This iterative process continues for subsequent rounds, where each round utilizes conditional probabilities with one additional dimension.

The comparison process remains largely unchanged for each round and does not vary significantly. Therefore, it is not reiterated here. However, it is important to note that as the rounds progress, deeper comparisons require increased computational resources. Furthermore, conducting additional rounds of comparison does not necessarily guarantee proportionate benefits. Thus, the number of rounds to be compared can be determined based on the specific communication needs and available computational resources.

Once the semantic information compression is performed by the BS, it transmits the conditions of each information compression to the user, along with the compressed triples. These compression conditions are represented as very short bit streams, which are insignificant compared to the size of the original data to be transmitted. Upon receiving the compressed information, the user can reconstruct the omitted information by leveraging the compressed conditions in conjunction with the common background knowledge $\mathbb{G}$.

\subsection{Joint Communication and Computation Resource Allocation}
The above reasoning process enables the exclusion of certain portions of the knowledge graph from transmission, thereby reducing communication delays. However, compressing information through reasoning requires additional time and energy for computation. To achieve improved results and minimize the overall system energy consumption during the communication process, it is crucial to consider the joint allocation of communication and computation resources. 

The total latency of the semantic communication system comprises two main components: the communication latency, denoted by $t_1$, and the computation latency, denoted by $t_2$. Note that the latency for processing and transmitting a text data message is constrained by a limit, denoted by $T$. Therefore, in order to meet the latency limit, the total process must adhere to the condition that the sum of communication latency and computation latency is less than or equal to $T$, i.e., $t_1 + t_2 \leq T$. This ensures that the overall latency of the system remains within the specified limit.

The communication time delay $t_1$ is calculated as follows.
The transmission rate between the BS and the user can be expressed as
\begin{equation}\label{eq15}
    c = B\log_2\left(1+\frac{ph}{\sigma^2}\right),
\end{equation}
where $B$ is the bandwidth of the system, $p$ is the transmit power of the BS, $h$ is the channel gain between the BS and the user, and $\sigma^2$ is the noise power.

If the head entity, relation, and tail entity in each triple are all encoded with the same number of bits $R$, then the total number of bits of semantic information $D$ in the text data can be calculated as
\begin{equation}\label{eq16}
    \operatorname{size}\left(D\right) = R\left(3M-E\right),
\end{equation}
where $M$ is the total number of triples in $D$, and $E$ is the number of triples in which the relation can be fully omitted.
Consequently, the communication latency can be written as
\begin{equation}\label{eq17}
    t_1 = \frac{\operatorname{size}\left(D\right)}{c} = \frac{R\left(3M-E\right)}{B\log_2\left(1+\frac{ph}{\sigma^2}\right)}.
\end{equation}

The computation time delay $t_2$ is calculated as follows.
The computational load of the proposed knowledge graph semantic information compression system is directly related to the number of comparisons in the compression process. To simplify the analysis, we will only consider the first and second rounds of comparison as described in the previous subsection. As it is infeasible to pre-determine the computational resources required to omit $E$ relations for a specific text data, we employ a statistical approach to calculate the ratio of the number of triples that can be omitted in each round to the total number of remaining triples. Based on this ratio, we estimate the computational resources required to omit $E$ relations in a statistical sense.

The ratio of the number of triples that can be omitted more in each round to the total number of triples before this round can be expressed as
\begin{equation}\label{eq18}
    Q = \left\{q_1,q_2,\cdots,q_n,\cdots,q_N\right\},
\end{equation}
where $q_1$ is the ratio of the number of triples that can be omitted more in the first round to the total number of all triples, $q_2$ is the ratio of the number of triples that can be omitted more in the first cycle of the second round to the number of triples whose relation is not omitted, and $q_n$ is the ratio of the number of triples that can be omitted more in the $\left(n-1\right)$-th cycle of the second round to the number of triples whose relation is not omitted before this cycle.

We can express the number of semantic relations that can be omitted in each round or cycle using a recursive formulation:
\begin{equation}\label{eq19}
    \left\{\begin{array}{l}
        E_1=M q_1, \\
        E_2=\left(M-E_1\right) q_2, \\
        E_3=\left(M-E_1-E_2\right) q_3, \\
        \vdots \\
        E_N=\left(M-E_1-E_2-\cdots-E_{N-1}\right) q_N,
    \end{array}\right.,
\end{equation}
where $M$ is the total number of triples, and $E_N$ is the number of relations which can be omitted in the $\left(N-1\right)$-th cycle of the second round.

Based on \eqref{eq19}, the computation load to omit $E$ relations in $M$ triples can be written as
\begin{equation}\label{eq20}
    l\left(E\right)=\left\{\begin{array}{l}
        \frac{E}{q_1}, 0 \leq E \leq E_1, \\
        \frac{E_1}{q_1}+\frac{\left(E-E_1\right) E_1}{q_2}, E_1<E \leq E_1+E_2, \\
        \frac{E_1}{q_1}+\left(\frac{E_1}{q_1}-E_1\right) E_1+\frac{\left(E-E_2-E_1\right) E_2}{q_3},\\ E_1+E_2<E \leq E_1+E_2+E_3, \\
        \vdots \\
        \frac{E_1}{q_1}+\left(\frac{E_1}{q_1}-E_1\right) E_1+\left(\frac{E_1}{q_1}-E_1-E_2\right) E_2+\\\cdots+\frac{\left(E-E_{N-1}-\cdots-E_1\right) E_{N-1}}{q_N},\\ \sum_{n=1}^{N-1} E_n<E \leq \sum_{n=1}^N E_n.
    \end{array}\right..
\end{equation}

\begin{figure*}[t]
\centering
\includegraphics[width=6.5in]{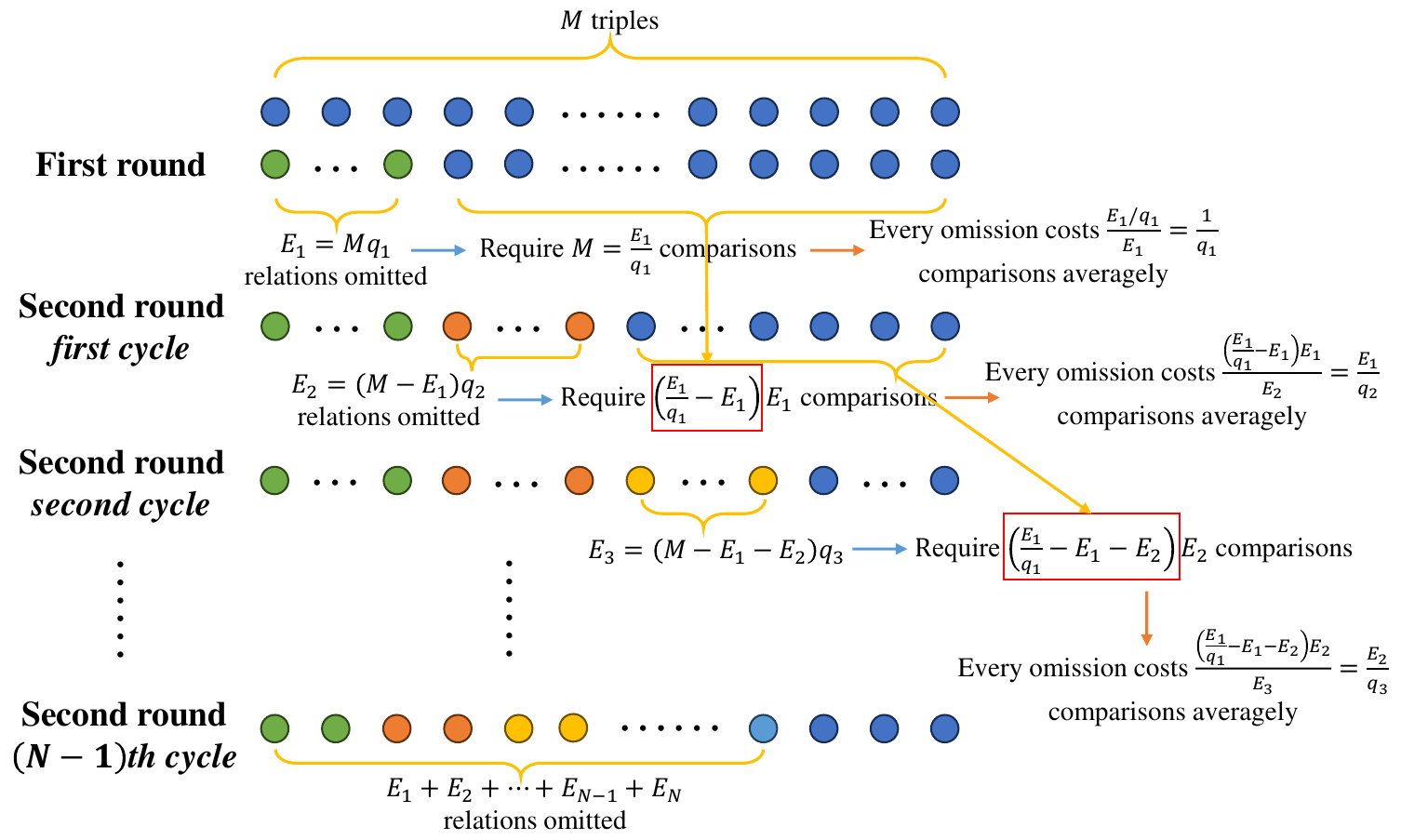}
\caption{Illustration of the insight of $l\left(E\right)$.}
\label{cl}
\end{figure*}

The piecewise function \eqref{eq20} of the computation load is derived from the number of comparisons required to omit $E$ relations. We calculate the average cost of one single omission at different round. The derivation process for $l\left(E\right)$ is shown in Fig. \ref{cl}.

It can be observed that $l\left(E\right)$ is an increasing segmented function, where each segment is a linear function with respect to $E$.
Consequently, the computation latency can be written as
\begin{equation}\label{eq21}
    t_2 = \frac{\tau_1 l\left(E\right)}{f},
\end{equation}
where $\tau_1$ is a constant coefficient, and $f$ is the computation capacity.

Based on the transmission time model, 
the communication energy consumption can be written as
\begin{equation}\label{eq22}
    e_1 = t_1 p.
\end{equation}
Besides, the computation energy consumption can be written as 
\begin{equation}\label{eq23}
    e_2 = \tau_1 \tau_2 l\left(E\right) f^2,
\end{equation}
where $\tau_1$ and $\tau_2$ are constant coefficients.

Based on the considered model, our aim is to minimize the total energy consumption of the semantic communication system, which can be formulated as
\begin{subequations}\label{eq24}
    \begin{align}
        \min_{p, E} \quad & e_1+e_2, \tag{\ref{eq24}}\\
        \textrm{s.t.} \quad & \frac{R(3 M-E)}{B \log _2\left(1+\frac{p h}{\sigma^2}\right)}+\frac{\tau_1 l\left(E\right)}{f} \leq T, \label{c1}\\
        & 0 \leq p \leq p_{\text{max}}, \label{c2}\\
        & 0 \leq E \leq M, E \in \mathbb{N} \label{c3}.
    \end{align}
\end{subequations}
Constraint (\ref{c1}) limits the total delay of communication and computation, while constraint (\ref{c2}) represents that the transmit power $p$ of the BS does not exceed a maximum limit $p_{\text{max}}$. Constraint (\ref{c3}) limits the total number of omitted relations $E$ to a non-negative integer that does not exceed the total number of triples $M$. The objective of the optimization problem is to minimize the total energy consumption of the semantic communication system.

\section{Algorithm Design}
In problem \eqref{eq24},
two variables need to be optimized: the transmit power $p$ of the BS and the total number of triples $E$ that omit semantic relations. 
Increasing the transmit power $p$ of the BS can reduce communication delay through increasing the transmission rate.
However, the increasing of power does not necessarily lead to reduced energy consumption in communication. 
Conversely, increasing the total number of triples $E$ that omit semantic relations can reduce communication delays, as fewer bits need to be transmitted. This can result in lower energy consumption in communication. However, it also increases computation delays, as more processing resource is required to compress the semantic information. 
The computation may consume more energy, offsetting the gains made in communication.

To minimize the total energy consumption of the system while satisfying all constraints, it is crucial to carefully select suitable values for the BS transmit power $p$ and the total number of triples $E$ that omit semantic relations. A balance needs to be struck, considering the specific requirements, constraints, and trade-offs involved in communication and computation energy consumption.

\subsection{Optimal Transmit Power}
To facilitate the analysis of the problem, we initially assume a fixed total number $E$ of triples that omit semantic relations and examine the influence of the BS transmit power $p$ on the overall system. We simplify constraint (\ref{c1}) and separate the transmit power $p$ of the BS to derive
\begin{equation}\label{eq25}
    p \geq\left(2^{\frac{R(3 M-E)}{B\left(T-\frac{\tau_1 l(E)}{f}\right)}}-1\right) \frac{h}{\sigma^2}.
\end{equation}
According to the right side of formula (\ref{eq25}), it can be observed that when $T-\frac{\tau_1 l(E)}{f}>0$, that is
\begin{equation}\label{eq26}
    l\left(E\right)<\frac{T \cdot f}{\tau_1},
\end{equation}
the right side of equation (\ref{eq25})
\begin{equation}\label{eq27}
    \left(2^{\frac{R(3 M-E)}{B\left(T-\frac{\tau_1 l(E)}{f}\right)}}-1\right) \frac{h}{\sigma^2}>0.
\end{equation}
When
\begin{equation}\label{eq28}
    \left(2^{\frac{R(3 M-E)}{B\left(T-\frac{\tau_1 l(E)}{f}\right)}}-1\right) \frac{h}{\sigma^2} \leq p_{\text{max}},
\end{equation}
we can combine constraint (\ref{c1}) with constraint (\ref{c2}) to get
\begin{equation}\label{eq29}
    \left(2^{\frac{R(3 M-E)}{B\left(T-\frac{\tau_1 l(E)}{f}\right)}}-1\right) \frac{h}{\sigma^2} \leq p \leq p_{\text{max}}.
\end{equation}

Expanding the objective function of the optimization problem, we can further obtain
\begin{equation}\label{eq30}
    F\left(p,E\right) = e_1+e_2 = \frac{R(3 M-E)p}{B \log _2\left(1+\frac{p h}{\sigma^2}\right)}+\tau_1 \tau_2 l\left(E\right) f^2.
\end{equation}
When $E$ is fixed, the objective function becomes a function with respect to $p$. We denote this function by $f\left(p\right)$, and we will analyze its monotonicity within the domain of definition (\ref{eq28}) as follows.

When $E$ is fixed, the second term of $f\left(p\right)$, which is $\tau_1 \tau_2 l\left(E\right) f^2$, is a constant. Therefore, we only need to analyze the monotonicity of the first term. To simplify this analysis, we can multiply the first term of $f\left(p\right)$ by the constant $\frac{h\cdot B}{\sigma^2 \cdot R\left(3M-E\right)}$. Let's denote this new expression as $g\left(p\right)$, and it can be written as
\begin{equation}\label{eq31}
    g\left(p\right) = \frac{\frac{p h}{\sigma^2}}{\log _2\left(1+\frac{p h}{\sigma^2}\right)}.
\end{equation}
It is evident that the monotonicity of $g\left(p\right)$ is the same as that of $f\left(p\right)$. Let's denote $\frac{ph}{\sigma^2}$ as $x$ and rewrite the new function as
\begin{equation}\label{eq32}
    h\left(x\right) = \frac{x}{\log _2\left(1+x\right)}.
\end{equation}
It is straightforward to observe that the monotonicity of the function $f(p)$ for $p > 0$ is equivalent to the monotonicity of the function $h(x)$ for $x > 0$.

Now, we analyze the monotonicity of the function $h\left(x\right)$.
First, taking the derivative of $h\left(x\right)$ with respect to $x$, we have
\begin{equation}\label{eq33}
    h^{\prime}(x)=\frac{\log _2(1+x)-\frac{x}{(\ln 2)(1+x)}}{\left[\log _2(1+x)\right]^2}.
\end{equation}
To analyze the monotonicity of $h(x)$, we need to find the sign of its derivative. Notice that the denominator of $h^{\prime}(x)$ is always positive when $x > 0$, so we only need to analyze the sign of its numerator.

Let
\begin{equation}\label{eq34}
    s(x)=\log _2(1+x)-\frac{x}{(\ln 2)(1+x)}.
\end{equation}
Then, the derivative of $s\left(x\right)$ with respect to $x$ is
\begin{equation}\label{eq35}
    s^{\prime}(x)=\frac{x}{(\ln 2)(1+x)^2}.
\end{equation}
Since $x>0$, $s^{\prime}(x)>0$, so $s(x)$ increases monotonically on $x>0$.
Due to the fact that 
\begin{equation}\label{eq36}
    s(0)=\log _2(1+0)-\frac{0}{(\ln 2)(1+0)}=0,
\end{equation}
we have $s(x)>0$ when $x>0$.
Therefore, when $x>0$, $h^{\prime}(x)>0$, the function $h(x)$ increases monotonically on the domain $x>0$.

It can be obtained that when $E$ is fixed, the objective function
\begin{equation}\label{eq37}
    f\left(p\right) = \frac{R(3 M-E)p}{B \log _2\left(1+\frac{p h}{\sigma^2}\right)}+\tau_1 \tau_2 l\left(E\right) f^2
\end{equation}
increases monotonically when $p>0$.

In summary, when the total number of triples that omit semantic relations $E$ is fixed, a smaller transmit power $p$ for the BS leads to better results, provided that the other constraints are satisfied. The optimal transmit power for the BS, satisfying constraints (\ref{eq26}) and (\ref{eq28}), is denoted as
\begin{equation}\label{eq38}
    p_{\text{optimized}} = \left(2^{\frac{R(3 M-E)}{B\left(T-\frac{\tau_1 l(E)}{f}\right)}}-1\right) \frac{h}{\sigma^2}.
\end{equation}

If constraint (\ref{eq26}) cannot be satisfied, it indicates that the value of $E$ is too large, and the computation delay alone exceeds the total latency limit. Therefore, this value of $E$ is undesirable. Additionally, since $l(E)$ is a monotonically increasing function, all values larger than this undesirable $E$ are also undesirable.
On the other hand, if constraint (\ref{eq26}) is satisfied but constraint (\ref{eq28}) cannot be satisfied, it means that the maximum transmit power of the BS becomes the limiting factor. In this case, the optimized transmit power $p_{\text{optimized}}$ is equal to the maximum transmit power $p_{\text{max}}$.

\subsection{Algorithm Analysis}
It follows that for each fixed value of $E$, we can determine the optimal value of the corresponding BS transmit power $p$. Given that the total number of triples that omit semantic relations, $E$, is a natural number and has an upper bound, it is advisable to design the algorithm based on traversing different values of $E$. The specific flow of the algorithm is outlined in Algorithm \ref{algo1}.

\begin{algorithm}[ht]
\algsetup{linenosize=\normalsize}
\normalsize
\caption{Joint Communication and Computation Optimization Algorithm for Text Semantic Compression}\label{algo1}
\begin{algorithmic}[1]
    \STATE Initialize $p_{\text{optimized}}=p_{\text{max}}$, $E_{\text{optimized}}=0$, and $obj_{\text{min}}=\inf$.
    \FOR{$E=0:M$}
        \IF{Constraint (\ref{eq26}) is not satisfied}
            \STATE Break
        \ENDIF
        \IF{Constraint (\ref{eq28}) is not satisfied}
            \STATE $p=p_{\text{max}}$
            \STATE $obj=F\left(p,E\right)$
        \ELSE
            \STATE Obtain $p$ according to equation (\ref{eq38})
            \STATE $obj=F\left(p,E\right)$
        \ENDIF
        \IF{$obj<obj_{\text{min}}$}
            \STATE $\left(obj_{\text{min}},p_{\text{optimized}},E_{\text{optimized}}\right)=\left(obj,p,E\right)$
        \ENDIF
    \ENDFOR
\end{algorithmic}
\end{algorithm}

According to Algorithm \ref{algo1}, the complexity of solving problem (\ref{eq24}) lies in the total number of triples to be transmitted. The total complexity of solving the joint communication and computation optimization problem is $\mathcal{O}(M)$, where $M$ is the total number of triples. 

To verify the correctness of the algorithm design, we utilized the official nonlinear programming solver \texttt{fmincon()} provided by MATLAB to solve this optimization problem. The results obtained from our algorithm were found to be consistent with the results obtained using \texttt{fmincon()}. Moreover, our algorithm demonstrated significantly faster speed compared to using \texttt{fmincon()}, thereby demonstrating the correctness and efficiency of our proposed algorithm.

\section{Simulation Results}
In the simulations, the maximum transmit power of the BS is 30 dBm. Unless specified otherwise, we set the bandwidth of the BS $B=10$ MHz, total number of triples $M=100$, and total latency limit $T=1$ ms. The values of most of the simulation parameters are set based on \cite{10032275}. The main system parameters are summarized in Table \ref{tb1}.

\begin{table}[ht]
\centering
\caption{Main System Parameters}
\begin{tabular}{cc}
    \toprule
    \textbf{Parameter}                                  & \textbf{Value} \\
    \midrule
    Bandwidth of the BS $B$                             & 10 MHz         \\
    Maximum transmit power of the BS $p_{\text{max}}$   & 30 dBm         \\
    Path loss $h$                                       & $10^{-9}$      \\
    Total number of triples $M$                         & 100            \\
    Total latency limit $T$                             & 1 ms           \\
    Computation capacity $f$                            & 1 GHz          \\
    \bottomrule
\end{tabular}
\label{tb1}
\end{table}

To evaluate the performance of the proposed joint communication and computation optimization system based on probability graph, labeled as `JCCPG', we designed two baseline algorithms for comparison. `Traditional' involves directly transmitting all triples with the same total delay, without considering the maximum transmission power limit. `Simplified JCCPG' optimizes the allocation of communication and computation resources based solely on the results of the first round of comparison.

\begin{figure}[t]
\centering
\includegraphics[width=3.2in]{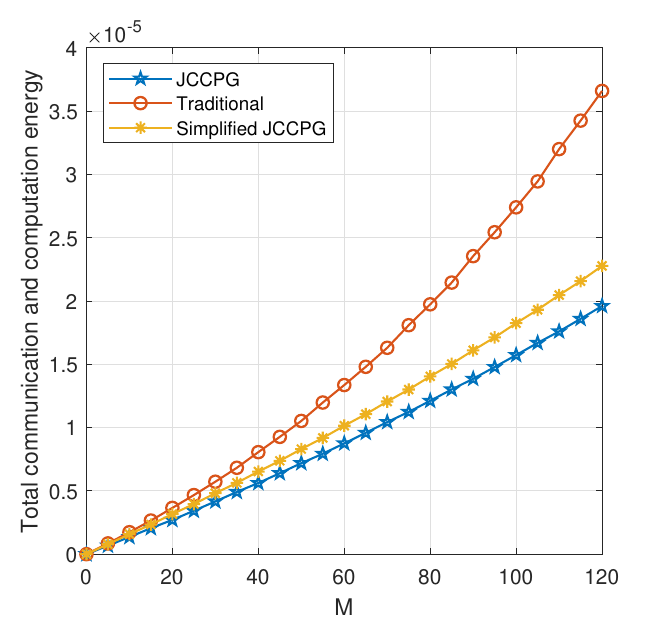}
\caption{Total communication and computation energy vs. total number of transmitted triples.}
\label{fig6}
\end{figure}

Fig. \ref{fig6} illustrates the variation trend of the total communication and computation energy consumption in the system with respect to the total number of transmitted triples. As the total number of transmitted triples increases, the energy consumption of all algorithms in the system also increases due to the increased transmission and computation requirements. However, `JCCPG' and `Simplified JCCPG' exhibit a significantly smaller growth rate compared to `Traditional'. This indicates the robustness of the proposed algorithm. Furthermore, the total system energy consumption of `JCCPG' is consistently lower than that of both `Traditional' and `Simplified JCCPG'. This advantage becomes more pronounced as the total number of transmitted triples increases. These findings highlight the significant advantage of the proposed algorithm in transmitting the large-scale data.

\begin{figure}[t]
\centering
\includegraphics[width=3.2in]{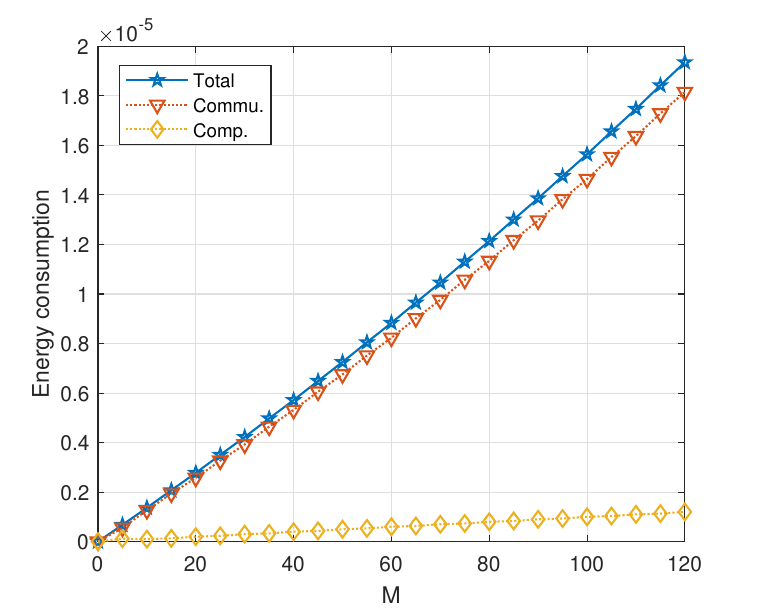}
\caption{Communication and computation energy consumption of `JCCPG' vs. total number of transmitted triples.}
\label{fig9}
\end{figure}

To gain a comprehensive understanding of the proposed `JCCPG' system, the energy consumption of `JCCPG' for communication and computation is illustrated in Fig. \ref{fig9}. The figure clearly demonstrates that the energy consumed by computation is significantly lower than that consumed by communication. This observation somehow suggests that computation exhibits greater energy efficiency compared to communication. Moreover, the future holds promise for considerable advancements in the computation capacity of terminal devices, accompanied by progressively more energy-efficient CPUs. Consequently, the integration of communication and computation is poised to become a prominent trend in the future as it possesses the potential to overcome the limitations of traditional communication methods.

\begin{figure}[t]
\centering
\includegraphics[width=3.2in]{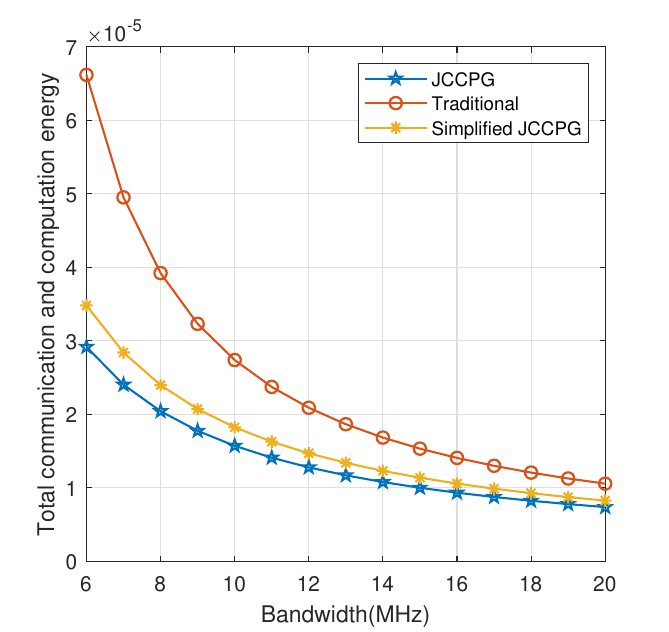}
\caption{Total communication and computation energy vs. bandwidth of the system.}
\label{fig7}
\end{figure}

Fig. \ref{fig7} illustrates the relationship between the total communication and computation energy consumption and the BS bandwidth. As the bandwidth of the BS increases, the total system energy consumption of all algorithms decreases. This is because a larger bandwidth provides more communication resources, resulting in shorter communication delay between the BS and the user. In `JCCPG', the reduced communication latency allows for longer computation latency, ultimately reducing the overall energy consumption of the system. When the bandwidth of the BS is large, the total energy consumption of the three algorithms tends to be similar. This is because when communication resources are abundant, the transmit power of the BS can be reduced. As a result, the communication energy consumption of all algorithms is small, and the proportion of computation becomes a relatively small factor. Therefore, the total energy consumption of the three algorithms does not differ significantly. However, when the bandwidth of the BS is small, indicating a poor communication environment, the proposed algorithm leverages the joint optimization of communication and computation, leading to a significant reduction in the total energy consumption of the system.

\begin{figure}[t]
\centering
\includegraphics[width=3.2in]{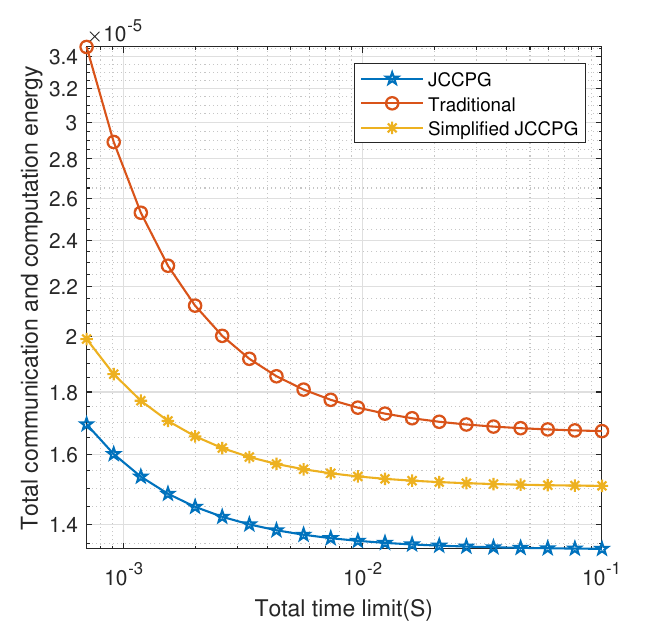}
\caption{Total communication and computation energy vs. total latency limit.}
\label{fig8}
\end{figure}

Fig. \ref{fig8} depicts the impact of the total delay constraint on the total communication and computation energy consumption of the system. When the total delay limit $T$ is small, `JCCPG' exhibits a significant advantage over `Traditional'. This is because in order to achieve the same total delay as `JCCPG', `Traditional' needs to substantially increase the transmit power of the BS, resulting in higher energy consumption for communication. In contrast, `JCCPG' reduces the amount of information to be transmitted through computation, thereby reducing the system's reliance on high transmit power and ultimately decreasing the total energy consumption. As the total delay limit $T$ increases to a certain value, the total system energy consumption of the three algorithms remains unchanged. This is because the total delay limit $T$ no longer acts as a constraint that impedes the reduction of total energy consumption in the system. At this point, the algorithms have optimized their communication and computation resources to achieve the minimum energy consumption within the given delay constraint.

In conclusion, the simulation results demonstrate that the proposed communication and computation joint optimization system, `JCCPG', achieves lower total communication and computation energy consumption compared to `Traditional' and `Simplified JCCPG'. Particularly in scenarios with poor communication conditions or when dealing with larger amounts of data to be transmitted, the proposed scheme exhibits significant advantages. These findings highlight the effectiveness of the proposed approach in reducing energy consumption and optimizing resource allocation in semantic communication systems.

\section{Future Work}
In the future, the model considered in this paper can be further developed. This section outlines some directions that could potentially enhance the model's capabilities.

Firstly, there's a prospect for dynamic refinement of the shared probability graph between the BS and the user during the communication process. Post transmission of semantic information for a given text data, the user could reconstruct an identical knowledge graph to the one possessed by the BS. By treating this knowledge graph as a new sample, both the BS and the user can collaboratively update the shared probability graph. The synchronicity in their approach ensures that the updated probability graph remains consistent. Consequently, there's no need for the BS to transmit the updated graph to the user after each update. The shared probability graph thus retains its single transmission requirement prior to formal text data transmission. Additionally, the notion of ``forgetting'' can be integrated into this update process. Essentially, more recent messages carry greater significance. This insight can be operationalized by assigning higher weight to recently transmitted knowledge graphs during probability graph updates. This approach ensures gradual retention of more recent, impactful information, aligning with the concept of age of information (AoI) \cite{yates2021age,kosta2017age}, a vibrant theme in wireless communication research.

Secondly, the current model doesn't harness the semantic correlations between distinct triples while calculating the multidimensional conditional probability. In practice, triples exhibit varying degrees of correlation, and leveraging the semantic information within the triples can enhance the efficiency of multidimensional conditional probability calculations. One possible way is to use existing large language models (LLMs) such as BERT \cite{devlin2018bert}, GPT \cite{lund2023chatting}, etc., to embed triples. This transforms the triples into vectors, with cosine similarity serving as the correlation coefficient. LLMs are extensively trained on copious amounts of high-quality data, showcasing their remarkable proficiency in processing textual information. Integrating LLMs into the domain of semantic communication stands as a potentially challenging yet promising research direction.

Thirdly, the present wireless communication model is a 
point-to-point configuration. 
Future endeavors could embrace more complex scenarios, like single-base station multi-user or multi-base station multi-user environments. This would entail devising optimization problems that better mirror real-world settings and exploring solutions in conjunction with communication techniques such as non-orthogonal multiple access (NOMA) \cite{saito2013non} and rate splitting multiple access (RSMA). 
These advancements would align more closely with actual communication dynamics and open up new dimensions of research.

\section{Conclusion}
In this paper, we have proposed a novel probability graph-based semantic information compression system designed for scenarios where the BS and the user share common background knowledge. By leveraging probability graph, we established a framework to represent and compress shared knowledge between the communicating parties. The proposed system operates by extracting semantic information from specific text data at the BS, which is then transformed into a knowledge graph. Using the shared probability graph, the BS selectively omits certain relation information during transmission. The compressed semantic data, along with the shared probability graph and predefined rules, enable the user to automatically restore the missing information upon reception. This approach effectively reduces communication resource consumption while introducing additional computational resource requirements. Considering the inherent limitations of wireless resources, we further addressed the challenge of joint communication and computation resource allocation. We formulated an optimization problem aiming at minimizing the total communication and computation energy consumption of the network, while satisfying constraints related to latency, transmit power, and semantic requirements. To solve this problem, we employed an efficient algorithm that demonstrates both correctness and efficiency. Simulation results validate the effectiveness of the proposed system. Specifically, the total communication and computation energy consumption of our system outperforms that of baseline algorithms in various scenarios. This highlights the significant advantages of our approach, particularly in scenarios with poor communication conditions or larger data volumes to be transmitted.

In conclusion, our work presented a comprehensive framework for probability graph-based semantic information compression, addressing resource optimization challenges in semantic communication systems. The proposed system showcases promising results, indicating its potential for improving communication efficiency, especially in challenging scenarios characterized by limited resources and substantial data volumes.

\bibliographystyle{IEEEtran}
\bibliography{main}

\end{document}